\documentclass[referee]{chjaa}          % referee version: for submission
\usepackage{graphicx,times}             %for PS/EPS graphics inclusion, new
\input{epsf.sty}                        %for PS/EPS graphics inclusion, old
\input{psfig.sty}                       %for PS/EPS graphics inclusion, old

\begin{document}

   \title{Observational constraints on quark matter in neutron stars
%\,$^*$
%\footnotetext{$*$ Supported by the National Natural Science Foundation of China.}
}
%   \subtitle{I. Place Your Subtitle Here}

   \volnopage{Vol.0 (200x) No.0, 000--000}      %%preserved for Editor. DOn't remove!
   \setcounter{page}{1}           %%starting page, preserved for Editor. DOn't remove!

   \author{Na-Na Pan
      \inst{1}\mailto{}
%% Please move "\mailto{}" to the corresponding author of the paper.
%% For single author or all the authors from an institute, use "\inst{}" only
%% Here is an example of three authors come from different institutes.
   \and Xiao-Ping Zheng
      \inst{1}
%%   \and
%%      \inst{3}
      }

   \institute{Institute of Astrophysics, Huazhong Normal University,
Wuhan 430079, P. R. China\\
             \email{Pannana@phy.ccnu.edu.cn}
%% Please give the E-mail address of the author, to whom future correspondence and
%% offprint requests will be sent. Note to pair \mailto{} with \email{}
%%        \and
%%              \\
%%        \and
%%             \\
          }

   \date{Received~~2001 month day; accepted~~2001~~month day}

   \abstract{
We estimate the constraints of observational mass and redshift on
the properties of equations of state (EOS) for quark matter based
on the quasiparticle description in the compact stars. We discuss
two scenarios: strange stars and hybrid stars. We construct the
equations of state utilizing an extended MIT bag model taking
medium effect into account for quark matter  and relativistic mean
field theory for hadron matter. We show that quark matter may
exist in strange stars and the interior of neutron stars. The bag
constant is a key parameter that affects strongly the mass of
strange stars. The medium effect can lead to the stiff hybrid-star
EOS directing to the pure hadronic EOS due to the reduction of
quark matter, and hence the existence of the heavy hybrid stars.
We find that the intermediate coupling constant may be the best
choice for compatibility with observational constraints in hybrid
stars.
   }

   \authorrunning{N.-N. Pan,  \& X.-P. Zheng }            %author_head in even pages
   \titlerunning{Observational constraints on quark matter in neutron stars }  % title_head in odd pages

   \maketitle
%% The author head (on even pages) and the title head (on odd pages) will be
%% automatically extracted from \author{} and \title{}. Whenever the title is too long,
%% you will be asked to supply a shorter one by inserting either \authorrunning{} or
%% \titlerunning{} before \maketitle. Anyway, you can specify your own heads in advance.
%%
%%
%% Note: In the following text body of your manuscript, please note several differences from
%%       other major journals:
%% (1) \subsection{Please Capitalize the First Letter of Each Notional Word in Subsection Title}
%% (2) Please Capitalize the First Letter of Each Notional Word in all tables' captions

%
%________________________________________________ sections below
%
\section{Introduction}           %% first-level sections will be auto-capitalized

The interiors of neutron stars contain matter at very high
densities that are a few times and even up to more than ten times
the density of ordinary atomic nuclei. This could provide a
high-pressure environment in which numerous subatomic particle
processes compete with each other. Wherefore, the components and
properties of interiors in neutron stars have attracted much
attention(Pandharipande \cite{1}; Glendenning \cite{2};
Glendenning et al. \cite{3};Sahu et al.\cite{4}; Kutschera \&
Koltroz \cite{5}; Thorsson et al. \cite{6}; Glendenning \cite{7};
Alford \& Reddy \cite{40}). But actually, the equation of state
for neutron star is still an indeterminacy in investigation for
many years on account of kinds of reasons. Observational
constraints on theoretical predictions of the equation of state of
high density matter have been habitually in practice(Glendenning
\cite{7}; Glendenning \& Moszkowski \cite{8}; Zhang et
al.\cite{49}).  Based on X-ray observations with the XMM Newton
observatory, Cottam et al. ({Cottam et al.\cite{27}} )reported
that neutron stars did not contain strange matter, but Xu ({Xu
\cite{48}}) addressed this conclusion was incorrect.  He found
that we still could not rule out strange star models for the X-ray
burster EXO 0748-676 from the mass-radius relations for bare and
crusted strange stars. Recently an accumulation of neutron-star
cooling observations also favors the presence of exotic particles
such as hyperons, quark matter in the cores of some neutron
stars(Yakovlev \& Pethick \cite{9}). Lackey et al. have
immediately given a particular discussion concerning the existence
of hyperons by comparing with the maximum observables, ie.,
stellar masses and gravitational redshifts(Lackey et al.
\cite{10}). Alford et al. and Kl\"{a}hn et al. have ever studied
 the effect of quark matter inside compact stars on the mass-radius
 relationship.
 Kl\"{a}hn et al. used $NJL(Nambu-Jona-Lasinio)$  model for quark matter (Kl\"{a}hn et al. \cite{42}).
Alford et al. modelled the quark matter equation of state through
a phenomenological parameterization (Alford et al. \cite{41}), but
their parameterization formalism was based on the consideration of
the perturbative QCD corrections. Moreover, they mainly focused on
the maximum masses of hybrid stars in accordance with the
construction of a sharp transition in their work. We also estimate
the circumstance with the inclusion of quark matter, but the first
transition in hybrid stars is based on the Gibbs construction. We
expect to uncover how the change of the region embodying mixed
phase and quark matter increases the maximum masses of hybrid
stars. We apply $GPS (Ghosh-Patak-Sahu)$ model (Ghosh et al.
\cite{16}) together with an extended MIT bag model for hybrid
stars. The extended MIT bag model is so-called "effective mass bag
model" presented by Schertler et al. (Schertler et al. \cite{17}).
In this model, medium effects are taken into account in the
framework of the MIT bag model by introducing density-dependent
effective quark masses. Such quark matter system in quasiparticle
description may involve partial nonperturbative contributions
since the coupling constant $g$ can be not small as shown below.
 According to the Tolman-Oppenheimer-Volkoff theory (Oppenheimer \& Volkoff \cite{18}),
differences of equations of state would cause different
mass-radius relations and in all the maximum masses. Based on mass
and radius of a neutron star, the gravitational redshift can
immediately be determined. Well then the observed neutron stars'
masses and gravitational redshifts would set a limit on the
equation of state. Our investigations emphasize on the influence
 of equation of state of quark matter on the limits of stars' mass and gravitational redshift
 under the consideration whether quark matter exist in compact
stars or not.

The most precise observations of neutron-star masses come from
radio pulsars in binaries, which are all measured with $95\%$
confidence to be less than 1.5 {$\textrm{M}_{\bigodot}$ (Thorsett
\& Chakrabarty \cite{19}), such as the most accurately measured
mass up to now but not necessarily the maximum possible mass of
PSR 1913+16 with $M=1.442\pm 0.003$ $\textrm{M}_{\bigodot}$
(Taylor \& Weisberg \cite{20}). X-ray measurements have long
suggested that accreting neutron stars are more massive, but the
contamination by oscillations of the high-mass main sequence
companion has been known (van Kerkwijk et al. \cite{21}). So the
record of 1.5 $\textrm{M}_{\bigodot}$ has remained the constraint
for many years until Nice et al. obtained a neutron-star mass
greater than 1.6 $\textrm{M}_{\bigodot}$ at the $95\%$ confidence
lever through recent radio observation of  PSR J0751+1807(Nice et
al. \cite{22}; Nice et al. \cite{23}; Nice et al. \cite{24}). Also
Ransom et al. discovered that the relativistic periastron advance
for the two eccentric systems in the globular cluster Terzan 5
indicates the existence of at least one of the pulsars that has
mass $>1.68$ $\textrm{M}_{\bigodot}$ at $95\%$ confidence(Ransom
et al. \cite{25}). Lackey et al. suggested that the  bound on the
maximum neutron-star mass is $\sim 1.68$ $\textrm{M}_{\bigodot}$
(Lackey et al. \cite{10}). However, $\ddot{O}$zel, in an analysis
of EXO 0748-676 observational data, found that the neutron-star
mass could only be low to 1.82 $\textrm{M}_{\bigodot}$ within
$1-\sigma$ bar ($\ddot{O}$zel \cite{26}). Although the result has
been recently disproved in an alternative analysis by Hynes et al.
(Hynes et al. \cite{47}), we still utilize the limit of mass
because the equations of state of pure hadron matter have
predicted such heavy stars . Another constraint is the measurement
of a gravitational redshift by Cottam et al.(Cottam et al.
\cite{27}). They have analysed the absorption lines in the spectra
of 28 bursts of the Low-mass X-ray binary EXO 0748-676. They
discovered that several absorbtion lines consistent with a
redshift $\emph{z}=0.35$ although with small uncertainties that no
more than $5 \%$ for the respective transitions.

The organization of the rest of this paper is as follows. In
section 2, we provide details of equations of state for strange
stars and hybrid stars. In section 3 and 4, we estimate the
constrains of observational mass and gravitational redshift on the
equations of state for strange stars and hybrid stars
respectively. Finally, we give the conclusions and discussions in
section 5.
%% Authors can use \cite, \citep and \citet for citation.
%% You may also give a citation as 'Michel et al. 1992', and use Table~1 or Fig.~1
%% and so forth. Using \ref and \label for cross-references of Tables/Figures is
%% a good way in adjusting/adding/removing text, tables or figures.

\section{Equation of state}

\subsection{ Strange stars}

A strange star is composed of pure strange quark matter, which is
made up of up, down, strange quarks and leptons. For  quark
matter, it is in nature that the equation of state must be
calculated by lattice quantum chromodynamics, but this couldn't be
carried through at finite density, so researchers often adopt some
phenomenological models in calculation(Baym \& Chin \cite{28};
Freedman \& Mclerran \cite{29}; Chakrabarty \cite{30}; Peng et al.
\cite{31}). Here we mainly take the effective mass bag model
considering medium effect into account(Schertler et al.
\cite{17}), which is based on quasiparticle approximation. We
could consider the u,d,s quarks to be quasiparticles  by utilizing
Debye screen effect in plasma. Under this circumstance, a quark
acquires an effective mass generated by the interaction with other
quarks of the dense system. The effective masses are derived from
the zero momentum limit of the dispersion relations following from
the quark self energy. In the hard dense loop (HDL) approximation,
the generic form of the quark self energy is
\begin{equation}
\Sigma =-a\> P_\mu \gamma ^\mu -b\> \gamma _0-c,
\end{equation}
thereinto
\begin{equation}
a=\frac{1}{4p^2}[tr(P_\mu \gamma ^\mu \Sigma )-p_0tr(\gamma
_0\Sigma )],
\end{equation}
\begin{equation}
b=\frac{1}{4p^2}[P^2tr(\gamma _0\Sigma )-p_0 tr(P_\mu \gamma^\mu
\Sigma )],
\end{equation}
\begin{equation}
c=-\frac{1}{4}tr\Sigma.
\end{equation}
Here $P^2=p_0^2-p^2$, then we can obtain the effective quark mass
\begin{equation}
m_{i}^{*}=\frac{m_{i}}{2}+\sqrt{\frac{m_{i}^{2}}{4}+\frac{g^{2}\mu_{i}^{2}}{6\pi^{2}}}.
\end{equation}
The mass formula depends on the coupling constant $g$, quark
chemical potential $\mu_{i}$ and the current mass of quark $m_{i}$
where $i=u,d,s$. In HDL approach, one expects that $g$ ought to be
small as effective mass is calculated perturbatively, but
Schertler et al. (Schertler et al. \cite{17}) extropolated it to
large value. We here take $g$ from 0 to 5 as done in previous
literatures (Schertler et al. \cite{17}; Schertler et al.
\cite{35}). The pressure and energy density for quark matter could
be constructed through the account of the statistical mechanics of
quasiparticles system,
\begin{equation}
\epsilon=\sum_{i}\{\frac{d}{16\pi^{2}}[\mu_{i}k_{i}(2\mu_{i}^{2}-m_{i}^{*2})-m_{i}^{*4}ln(\frac{k_{i}+\mu_{i}}{m_{i}^{*}})]+B^{*}(\mu_{i})\}+\epsilon_{e}+B,
\end{equation}
\begin{equation}
p=\sum_{i}\{\frac{d}{48\pi^{2}}[\mu_{i}k_{i}(2\mu_{i}^{2}-5m_{i}^{*2})+3m_{i}^{*4}ln(\frac{k_{i}+\mu_{i}}{m_{i}^{*}})]-B^{*}(\mu_{i})\}+p_{e}-B,
\end{equation}
where $d$ is the degree of degeneracy and $B^{*}$ is a function to
maintain thermodynamic self-consistency,
\begin{equation}
  \frac{dB^{*}(\mu_{i})}{dm_{i}^{*}}=-\frac{d}{4\pi^{2}}\left[m_{i}^{*}\mu_{i} k_{i}-
  m_{i}^{*3} \ln (\frac{k_{i}+\mu_{i}}{m_{i}^{*}})\right].
\end{equation}
Up to now, it is difficult to estimate bag constant $B$ and
strange quark mass $m_{s}$ from available data. They are subject
to systematic uncertainties, so we typically treat them as the
free parameters ranging from $140^{4}$ to  $200^{4}$
$\textrm{Mev}^{4}$ for $B$ and from $80$ to $150$ $\textrm{Mev}$
for $m_{s}$ as many researchers have done.

\subsection{ Hybrid stars}

A hybrid star mainly contains quark matter core, mixed
quark-hadron phase and hadron matter if the surface tension at the
boundary between the quark and hadron phase is low enough.
 To construct the equation of state for hybrid stars, we  are firstly in need of  the equations of state for hadron
 matter and quark matter. The equation of state of quark matter has been discussed in previous section. Generally speaking,
we should use the $Baym-Pethick-Sutherland$ ($BPS$)(Baym et al.
\cite{32}) equation of state for subnuclear densities
corresponding to the crust of the star, which is matched with the
equation of state for nuclear densities at
$\epsilon\approx10^{14}\textrm{g/cm}^{3}$. The equation of state
of hadron matter for nuclear densities could be established in
relativistic mean field theory(Glendenning \cite{7}) that is one
of the effective field theories describing hadron matter, where
nucleons interact through the nuclear force mediated by the
exchange of isoscalar and isovector mesons ($\sigma, \omega,
\rho$).  We here consider the hadron phase including only  four
kinds of particles n, p, e, $\mu$.  A relativistic Lagrangian
reads
\begin{eqnarray}\label{l1}
 \textit{ \L}=\sum_{B=n,p}\overline{\psi}_{B}(i\gamma_{\mu}\partial^{\mu}-m_{B}+g_{\sigma B}\sigma-g_{\omega B}\gamma_{\mu}\omega^
{\mu}-\frac{1}{2}g_{\rho
B}\gamma_{\mu}\tau\cdot\rho^{\mu})\psi_{B}\nonumber \\
+\frac{1}{2}(\partial_{\mu}\sigma\partial^{\mu}\sigma-m_{\sigma}^{2}\sigma^{2})-U(\sigma)+
\sum_{\lambda=e,\mu}\overline{\psi}_{\lambda}(i\gamma_{\mu}\partial^{\mu}-m_{\lambda})\psi_{\lambda}\nonumber \\
-\frac{1}{4}\omega_{\mu\nu}\omega^{\mu\nu}+\frac{1}{2}m_{\omega}^{2}\omega_{\mu}\omega^{\mu}-
\frac{1}{4}\rho_{\mu\nu}\cdot\rho^{\mu\nu}\nonumber \\
+\frac{1}{2}m_{\rho}^{2}\rho_{\mu}\cdot\rho^{\mu} .
\end{eqnarray}
We then solve the Euler-Lagrange equations of (\ref{l1}) by
replacing the fields by their mean values under the assumption
that the bulk matter is static and homogeneous and then calculate
the kinetic Dirac equations for baryons and also the meson fields
equations. By imposing $\beta$-equilibrium, local electric charge
neutrality and conservation of baryon number, we obtain the
equation of state for pure hadron matter,
\begin{eqnarray}
  \epsilon=\frac{1}{3}bm_{n}(g_{\sigma}\sigma)^{3}+\frac{1}{4}c(g_{\sigma}\sigma)^{4}
  +\frac{1}{2}m_{\sigma}^{2}\sigma^{2}+\frac{1}{2}m_{\omega}^{2}\omega_{0}^{2}
  +\frac{1}{2}m_{\rho}^{2}\rho_{03}^{2}\nonumber \\
  +\sum_{B=n,p}\frac{2J_{B}+1}{2\pi^{2}}\int_{0}^{k_{B}}\sqrt{k^{2}+(m_{B}-g_{\sigma B}\sigma)^{2}}
  k^{2}dk \nonumber \\
  +\sum_{\lambda}\frac{1}{\pi^{2}}\int_{0}^{k_{\lambda}}\sqrt{k^{2}+m_{\lambda}^{2}}k^{2}dk,
\end{eqnarray}
\begin{eqnarray}
 p=-\frac{1}{3}bm_{n}(g_{\sigma}\sigma)^{3}-\frac{1}{4}c(g_{\sigma}\sigma)^{4}
  -\frac{1}{2}m_{\sigma}^{2}\sigma^{2}+\frac{1}{2}m_{\omega}^{2}\omega_{0}^{2}
  +\frac{1}{2}m_{\rho}^{2}\rho_{03}^{2}\nonumber \\
  +\frac{1}{3}\sum_{B=n,p}\frac{2J_{B}+1}{2\pi^{2}}\int_{0}^{k_{B}}\frac{k^{4}}{\sqrt{k^{2}+(m_{B}-g_{\sigma B}\sigma)^{2}}}
 dk \nonumber \\
 +\frac{1}{3}\sum_{\lambda}\frac{1}{\pi^{2}}\int_{0}^{k_{\lambda}}\frac{k^{4}}{\sqrt{k^{2}+m_{\lambda}^{2}}}dk .
\end{eqnarray}
Actually, the Euler-Lagrange  equations feature five free
parameters, which under certain assumptions are fit algebraically
to numbers distilled from laboratory measurements of many finite
nuclei, ie., the saturation density, binding energy per nucleon
and symmetry energy coefficient at saturated nuclear matter, and
the overall compressibility  $K$ and the effective mass of
nucleons $m^{*}$.  As it is not our motive here, we numerically
adopt five fiducial equations of state using $Glendenning$ ($GL$)
(Glendenning \cite{7}) and $Ghosh-Patak-Sahu$ ($GPS$) (Ghosh et
al. \cite{16}) parameters for demonstration, which are arranged in
Table 1 and represent the equation of state from the softest to
the stiffest respectively.

Under the consideration of the models of the compact stars inside
which the  deconfinement transition occurs at high density
(Glendenning \cite{33}; Glendenning  \cite{7}), we allow hadron
phase to undergo a first order phase transition (Schertler et al.
\cite{34}; Schertler et al. \cite{35})to a
 deconfined quark matter phase above the saturation density of nucleon. This phase transition makes it
 possible that the occurrence of a mixed hadron-quark phase in a
 finite density range inside compact stars. Based on the quark and hadron matter equations of state,  $\beta$-equilibrium, global
electric charge neutrality and Gibbs condition between quark and
hadron phases, we could easily get the equation of state for mixed
phase, and then the equation of state for hybrid stars.

\section{Maximum mass}
The structure of a neutron star is determined by the local balance
between the attractive gravitational force and the pressure force
of the neutron star matter. For an equilibrium neutron star under
the condition of overlooking the effect of rotation, the
gravitational field is taken to be static and spherically
symmetric. Considering the effect of general relativity, Tolman,
Oppenheimer and Volkoff established a set of equations (TOV
equations) (Oppenheimer \& Volkoff \cite{18}) to determine the
structures of these stars:
\begin{equation}
\frac{dp(r)}{dr}=-\frac{[\epsilon(r)+p(r)][m(r)+4 \pi
r^{3}p(r)]}{r[r-2m(r)]},
\end{equation}
\begin{equation}
\frac{dm(r)}{dr}=4\pi r^{2}\epsilon(r).
\end{equation}
Here $G=c=1$, $p(r)$ and $\epsilon(r)$ are the pressure and energy
density of the matter at the radius $r$, and $m(r)$ is the total
mass inside the star within a sphere of given radius $r$:
\begin{equation}
m(r)=4\pi \int_{0}^{r}\epsilon(r')r'^{2}dr'.
\end{equation}
After the equation of state of the star
$\epsilon(r)=\epsilon(p(r))$ is given, the TOV equations could be
finally solved as an initial value problem. Stating the values of
the pressure $p(r=0)=p_{c}$ and mass $m(r=0)=0$ in the center of
the star, we can integrate the TOV equations outwardly until the
surface $p(r=R)=0$ is reached, then we could give the $M-R$
relation of the star, which has a maximum mass under the
prediction of the general relativity. Using this property, we can
rule out those equations of state that are too soft to produce the
observed masses. Actually, Ter 5 I rotates at 104 \textrm{Hz}
(Ransom \cite{25})and EXO 0748-676 at 45 \textrm{Hz} (Villarreal
\& Strohmayer \cite{36}), which may increase the TOV maximum mass
less than $1\%$. On analysing the theoretical results with
observational data, we should also take this small correction into
account.

In Fig.1, we plot the $R-M$ relation for strange stars based on
our fiducial effective mass bag model considering medium effect.
We find that the consideration of medium effect and the increase
of the current mass of s quark $m_{s}$ could only slightly soften
the equation of state, but the change of bag constant $B$ can
affect the stiffness of equation of state distinctly. And only the
ones with smaller bag constant are necessary for strange stars to
be consistent with the higher mass limits.

We investigate the influence of the hadronic equations of state
 on
the ones for hybrid stars in Fig.2. It displays little effect on
the stiffness of the equations of state for hybrid stars and in
all the maximum mass, which could also be seen below for
gravitational redshift. Our choice of the intermediate one,
$GPS2$, as our primary to estimate the equations of state of
hybrid stars suffices present needs. In Fig.3, we  show the $R-M$
relation for hybrid stars. The  symbol crosses on each curve
represent the dividing point of the hadron phase and mixed phase.
Only the compact stars lie in the district between the cross and
the maximum mass point could have quark matter in the interior.
The coupling constant for strong action $g$ as well as the bag
constant $B$ influence the stiffness of the equation of state for
hybrid stars remarkably. For $g=0.0$ condition, almost all the
equations of state with quark matter  are ruled out by both limits
of mass, except the stiffest one with parameters $B^{1/4}=200.0
\textrm{Mev}, m_{s}=150.0 \textrm{Mev}$ which can be brought
within the $95\%$ confidence limit of 1.68 $\textrm{M}_{\bigodot}$
from Ter 5 I account for the rotation correction. Under $g=3.0$,
both curves are consistent with the limit from Ter 5 I but only
the ones with $B^{1/4}=200.0 \textrm{Mev}$ could fit the 1.82
$\textrm{M}_{\bigodot}$. And when $g=4.0$, all equations of state
reach the larger constraint 1.82 $\textrm{M}_{\bigodot}$.  For
$g=5.0$, quark matter disappears in the neutron stars. The
equations of state of pure nucleonic matter are consistent with
1.68 $\textrm{M}_{\bigodot}$ and 1.82 $\textrm{M}_{\bigodot}$.
Obviously, the hybrid stars with medium effect of quark matter are
very different from the ones without medium effect. The reason for
the existence of heavy hybrid star is that increasing $g$ and $B$
soften the EOS of quark matter, which heightens the transition
density and cause difficulty to occur from hadrons to quark matter
and in all hardens the EOS of hybrid star. This effect brings
about a hybrid star dominated by hadronic matter, i.e., the extent
of quark matter reduces but the hadron matter region increases and
this in fact yields "hybrid" stars that are actually hadronic
stars with a tiny core of mixed quark-hadron matter inside and the
most of the mass and radius contributions come from the hadron
matter, so the Mass-radius relation of hybrid star has the
tendency towards the pure neutron star and the heavier stars
appear.

\section{Gravitational redshift}
As is known by deduction of general relativity, when photons emit
from radiant point in a gravitational field, the spectrum line
would shift to the red part of spectrum observed from the place
apart from the field. This phenomena is called gravitational
redshift. Generally speaking, the quantity of gravitational
redshift is small, expect around the black holes or neutron stars
which have a strong gravitational field. For a nonrotating star,
the redshift $\emph{z}$ obeys the relation
\begin{equation}
\emph{z}=(1-\frac{2M}{R})^{-1/2}-1,
\end{equation}
which is evidently relevant to the value of $M/R$ and could permit
a determination of the mass-to-radius ratio as its potential
usefulness in measurements for compact stars. Attentively, we have
already known from the curve of $M-R$ relation that $R$ decreases
with $M$ until $M$ approaches its maximum for a stable star. The
redshift spontaneously has a maximum at the point of maximum-mass
star, which could be also used to rule out the equations of state
that can't produce a redshift in observation. Cottam et al.
(Cottam et al. \cite{27})have analysed the absorption lines in the
spectra of 28 bursts of the Low-mass X-ray binary EXO 0748-676.
They identified the most significant features with Fe XXVI and XXV
$n=2-3$ and O VIII $n=1-2$ transitions and discovered all
transitions with a redshift $\emph{z}=0.35$ although with small
uncertainties for the respective transitions. Recent observation
suggests that the neutron star rotates at 45 \textrm{Hz}
(Villarreal \& Strohmayer \cite{36}), but we still consider
nonrotating approximation as the rotation correction to the
redshift should be small.

We plot the redshift as a function of mass for strange stars based
on our fiducial effective mass bag model considering medium effect
in Fig.4. All the equations of state are consistent with
$\emph{z}=0.35$. Together with the maximum mass constraint
discussed in the section 3, we conclude that  the observation mass
constraint  acts on the equations of state for quark matter
strongly, as well as the free parameters especially the bag
constant $B$ in strange stars.

Analogically, we investigate the influence of the hadronic
equations of state under the consideration of relativistic
mean-field theory on $Redshift(M)$ of hybrid stars in Fig.5 as
done in Fig.2. The maximum reshifts are not sensitive to the
equations of state for hadrons but to the ones of  quark matter at
several times nuclear density, which could also be seen in Fig.6.
Contrast to the maximum masses limits,   the ones with
$B^{1/4}=170.0 \textrm{Mev}$ under $g=0.0$ are both consistent
with the redshift constraint 0.35, and the circs for
$B^{1/4}=200.0 \textrm{Mev}$ are opposite. This means that both
equations of state without medium effect must be excluded by the
observational constraints. For $g=3.0$ and $g=4.0$, all equations
of state with quark matter are consistent with $\emph{z}=0.35$.
The equations of state under the case $g=5.0$ are also compatible
with the constraint but quark matter is absent.

\section{Conclusion}
We have compared equations of state for quark matter in
quasiparticle description in strange stars and hybrid stars with
astronomical observations of mass and gravitational redshift. We
study effect of the coupling strength among quarks on strange
stars and hybrid stars. In strange stars, the effect of the
coupling constant is small to reach the observational limits. In
hybrid stars, however, the coupling constant play an important
role in maximum masses and gravitational redshifts of the stars.
We think that hybrid star is a more realistic model, where
intermediate dense strange quark matter is possible. The hybrid
stars have masses much lower than the observed masses when quark
matter exists as a free fermion gas, but can produce the observed
mass and redshift  when medium effect of strange quark matter is
taken into account. The studies in section 3 and 4 have shown that
intermediate coupling constant may be the best choice for making
the equations of state for hybrid stars consistent with both the
observation mass and redshift constraints better for a wide range
of parameter $B$.

In quality, we can obtain the similar maximum masses given by
Alford et al.(Alford et al. \cite{41}) that the presence of quark
matter in the neutron stars is constrained but not ruled out by
the observational mass and gravitational redshift. However, there
are quite evident differences between the investigation of Alford
et al. and ours. Firstly, our model contains large coupling
constant $g$ (or $\alpha_{c}=g^{2}/4\pi^{2}
>1$) which is slightly different from parameterized EOS  based on
perturbative QCD theory. Secondly, the heavier hybrid star is
dominated by quark matter for the case of the sharp transition
while in our configuration by hadronic matter.  The reason is that
the softening of EOS of  quark matter leads to the existence of
the almost pure hadron star in our model. In sharp transition,
Alford et al.  owe it to the hardening of the quark matter EOS by
increasing c and in all the existence of a much more larger quark
matter core. Our work also indicates that there is a critical
value of $g\sim5.0$ over which quark matter connot exists inside
the stable neutron stars.

It is well-known that cooling measurements is an additional
constraints on equation of state of compact stars. Current
measured temperatures of compact objects are compatible with
strange stars and hybrid stars (Yu \& Zheng \cite{37}; Kang \&
Zheng \cite{38}). Existence of hyperons and quark matter in
compact objects indicates that these such equations of state tend
to result in excessively fast cooling due to the onset of direct
Urca processes, and so is the stiff nucleonic equation of state in
heavier stars(Page et al. \cite{39}). Thus the normal neutron
stars or hyperon stars may be much cooler to be ruled out by X-ray
data. However, the  hybrid stars do not have this problem because
a deconfinement heating mechanism is triggered due to the
spin-down of the stars. We discovered that the deconfinement
latent heat can effectively cancel the enhanced neutrino emission
(Kang \& Zheng \cite{38}).

\begin{acknowledgements}
This work is supported by NSFC under Grant Nos. 10373007 and
10603002.
\end{acknowledgements}

\clearpage

\begin{table*}
\begin{center}
\begin{tabular}{|c|c|c|c|c|c|c|}\hline
$Name$ & $\rho_{0}(\textrm{fm}^{-3})$ & $B/A(\textrm{Mev})$ &
$a_{sym}(\textrm{Mev})$& $K(\textrm{Mev})$ & $m^{*}/m_{N}$ & $EOS$
\\ \hline
$GL1$ & 0.153 & -16.3 & 32.5 & 240 & 0.78 & softest \\
\hline $GPS1$ & 0.150 & -16.0 & 32.5 & 250 & 0.83 & soft \\ \hline
$GPS2$ & 0.150 & -16.0 & 32.5 & 300 & 0.83 & intermediate \\
\hline $GPS3$ & 0.150 & -16.0 & 32.5 & 350 & 0.83 & stiff \\
\hline $GL2$ & 0.153 & -16.3 & 32.5 & 300 & 0.7 & stiffest
\\\hline
\end{tabular}
\end{center}
\vspace{5mm} \caption{Parameters for five fiducial hadronic
equations of state in relativistic mean-field theory. $\rho_{0}$
is the saturation density, $B/A$ is the binding energy, the
incompressibility is denoted by $K$, the effective mass by
$m^{*}/m_{N}$ and the symmetry energy by $a_{sym}$. }.
\label{parameter}
\end{table*}
\begin{figure}
\includegraphics[width=0.9\textwidth]{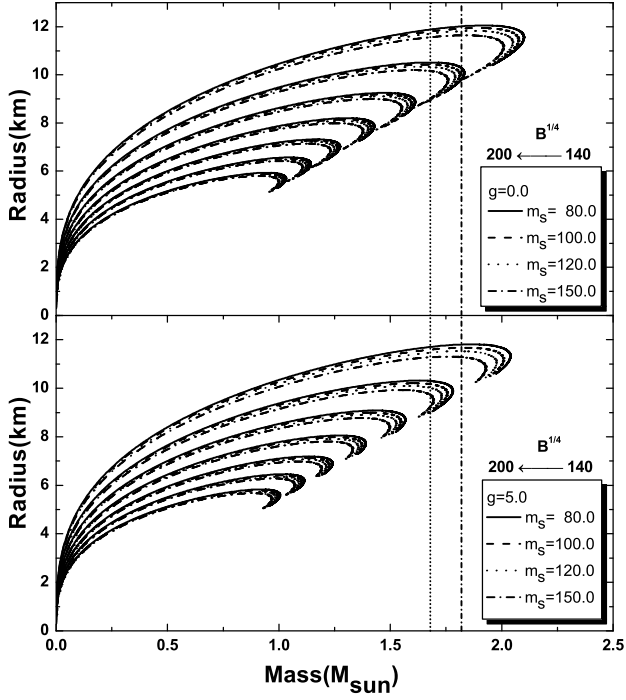}
\caption{Oppenheimer-Volkoff radius-mass curves for strange stars
using effective mass bag model considering medium effect (bottom)
and not (top). These two vertical lines represent the
observational $95 \%$ confidence limit 1.68
$\textrm{M}_{\bigodot}$ from Ter 5 I and 1.82
$\textrm{M}_{\bigodot}$ within $1-\sigma$ bar from EXO 0748-676.
$B$, $g$ and $m_{s}$ are bag constant, coupling constant and the
current mass of s quark respectively. These seven sets of four
curves in each panel represent the results assuming seven
different values about $B^{1/4}$ ranging from 200 \textrm{Mev}
(ie. the left set) to 140 \textrm{Mev} (ie. the right set) with
equal intervals of 10 \textrm{Mev}.} \label{strange star mr}
\end{figure}
\begin{figure}
\includegraphics[width=0.9\textwidth]{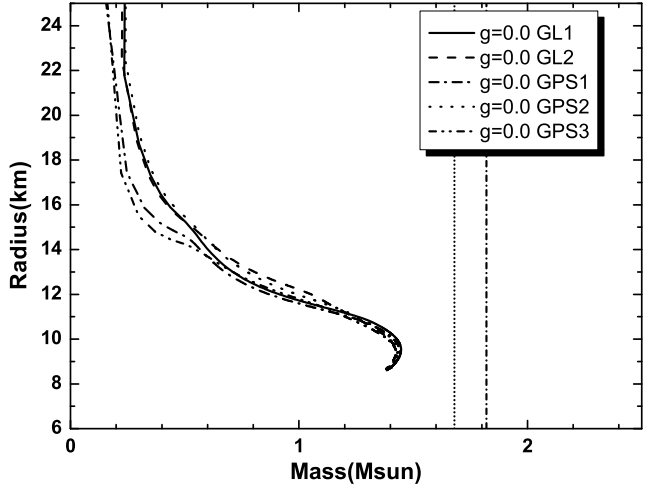}
\caption{Oppenheimer-Volkoff radius-mass curves for hybrid stars
with the same equation of state for quark matter and different
equation of state for hadrons ($GL1, GPS1, GPS2, GPS3, GL2$).
These two vertical lines represent the observational $95 \%$
confidence limit 1.68 $\textrm{M}_{\bigodot}$ from Ter 5 I and
1.82 $\textrm{M}_{\bigodot}$ within $1-\sigma$ bar from EXO
0748-676. For quark matter, we choose the parameters
$B^{1/4}=170.0  \textrm{Mev}$, $g=0.0$ and $m_{s}=150.0
 \textrm{Mev}$ in calculation.} \label{comparison}
\end{figure}
\begin{figure}
\includegraphics[width=0.9\textwidth]{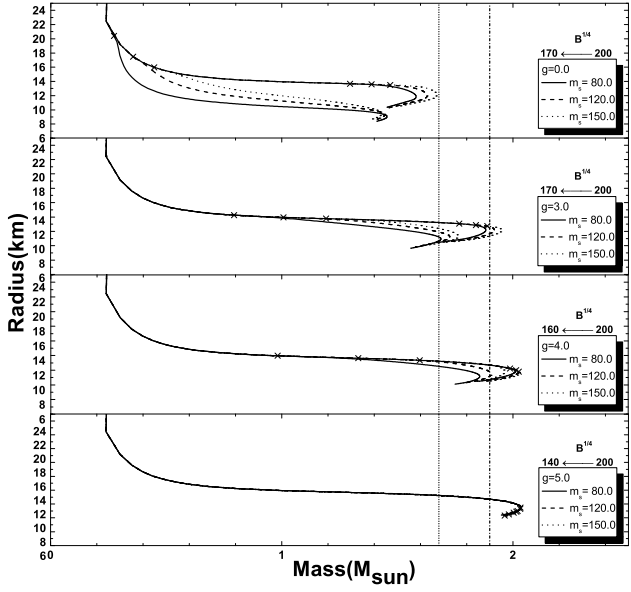}
\caption{Oppenheimer-Volkoff radius-mass curves for hybrid stars
with same equation of state for hadrons ($GPS2$), and the symbol
cross on each curve represents the dividing point of the hadron
phase and the mixed phase. These two vertical lines represent the
observational $95 \%$ confidence limit 1.68
$\textrm{M}_{\bigodot}$ from Ter 5 I  and 1.82
$\textrm{M}_{\bigodot}$ within $1-\sigma$ bar from EXO 0748-676.
These two sets of three curves represent the results of two
different values about $B^{1/4}$ which ranges from 170
\textrm{Mev} (ie. the left set) to 200  \textrm{Mev} (ie. the
right set) in the upper two panels, from 160 \textrm{Mev} (ie. the
left set) to 200 \textrm{Mev} (ie. the right set) in the third
panel, and from 140  \textrm{Mev} (ie. the left set) to 200
\textrm{Mev} (ie. the right set) in the fourth panel.}
\label{hybrid star mr}
\end{figure}
\begin{figure}
\includegraphics[width=0.9\textwidth]{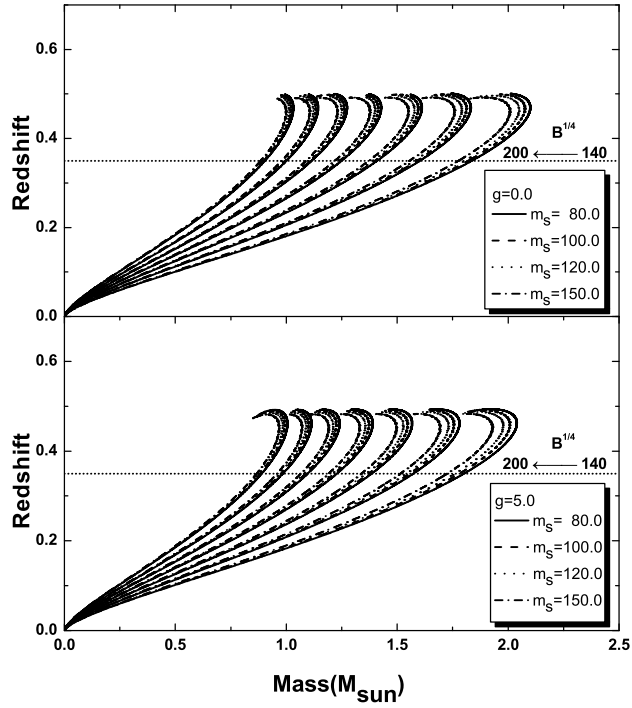}
\caption{Gravitational redshift vs mass for strange stars using
effective mass bag model considering medium effect (bottom) or not
(top). The horizontal line is $\emph{z}=0.35$ measured for
EXO0748-676. These seven sets of four curves in each panel
represent the results assuming seven different values about
$B^{1/4}$ ranging from 200 \textrm{Mev} (ie. the left set) to 140
\textrm{Mev} (ie. the right set) with equal intervals of 10
\textrm{Mev}.} \label{strange star mredshift}
\end{figure}

\begin{figure}
\includegraphics[width=0.9\textwidth]{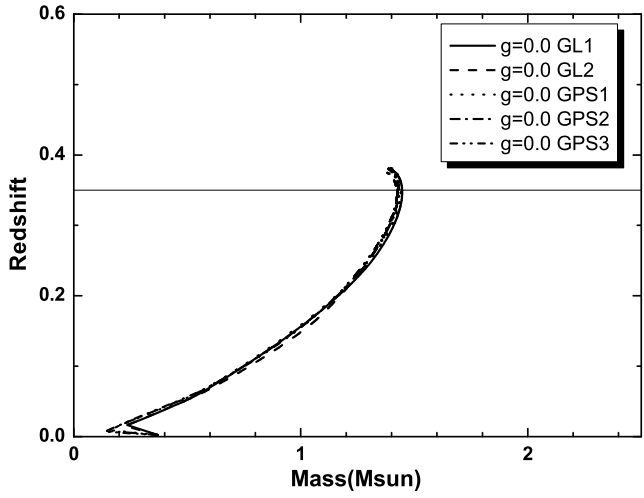}
\caption{Gravitational redshift vs mass for hybrid stars with the
same equation of state for quark matter and different equation of
state for hadrons ($GL1, GPS1, GPS2, GPS3, GL2$). The horizontal
line is $\emph{z}=0.35$ measured for EXO0748-676. For quark
matter, we choose the parameters $B^{1/4}=170.0  \textrm{Mev}$,
$g=0.0$ and $m_{s}=150.0  \textrm{Mev}$ in calculation.}
\label{comparisonmredshift}
\end{figure}
\begin{figure}
\includegraphics[width=0.9\textwidth]{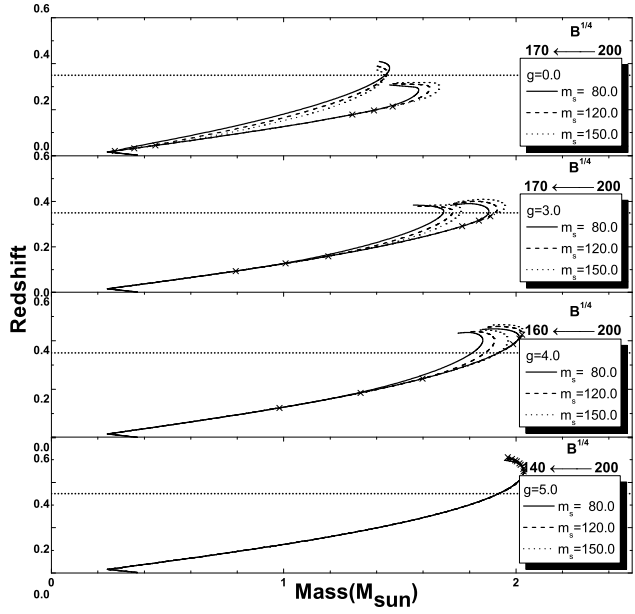}
\caption{Gravitational redshift vs mass for hybrid stars with same
equation of state for hadrons ($GPS2$), and the symbol cross on
each curve represents the dividing point of the hadron phase and
the mixed phase. The horizontal line is $\emph{z}=0.35$ measured
for EXO0748-676.  These two sets of three curves represent the
results of two different values about $B^{1/4}$ which ranges from
170 \textrm{Mev} (ie. the left set) to 200  \textrm{Mev} (ie. the
right set) in the upper two panels, from 160 \textrm{Mev} (ie. the
left set) to 200 \textrm{Mev} (ie. the right set) in the third
panel, and from 140  \textrm{Mev} (ie. the left set) to 200
\textrm{Mev} (ie. the right set) in the fourth panel.}
\label{hybrid star mredshift}
\end{figure}

\label{lastpage}

\end{document}